\documentclass[twocolumn,showpacs,prb,amsmath,amssymb]{revtex4}

\DeclareMathOperator{\tRe}{Re}
\DeclareMathOperator{\tIm}{Im}

\usepackage{graphicx}
\usepackage{bm}
\usepackage[varg]{txfonts}

\begin{document}

\title{Universal temperature corrections to  Fermi liquid theory in
  an interacting electron system}

\author{V. M. Galitski}
\author{S. \surname{Das Sarma}}
\affiliation{Condensed Matter Theory Center,
Department of Physics, University of Maryland, College Park, Maryland
20742-4111}

\begin{abstract}
We calculate analytically the effective mass and the quasiparticle
renormalization factor in an electron liquid with long-range Coulomb
interactions between electrons in two and three dimensions in the
leading order density expansion. We concentrate on the temperature
dependence of the effective mass in the limit $T/T_{\rm F} \ll r_{\rm
  s} \ll 1$ and show that the leading temperature correction is linear
in two dimensions and proportional to $T^2 \ln \left( 1/T\right)$ in
three dimensions (positive in both cases).  We explicitly calculate
the coefficients, which are shown to be universal density independent
parameters of the order of unity (in the high-density limit). The
singular temperature corrections are due to the singularity in the
dynamic dielectric function at $\omega \sim v_{\rm F} q$ and $q \ll 2
p_{\rm F}$.  In two dimensions, we predict a non-monotonic effective
mass temperature dependence and find that the maximum occurs at a
temperature $T^* \sim T_{\rm F} r_{\rm s} \ln^{-1}\left({1/r_{\rm
    s}}\right)$. We also study the quasiparticle renormalization
factor in both three and two dimensions.
\end{abstract}


\maketitle


\section{Introduction}
The basic postulate of  Fermi liquid theory is the existence of a
one-to-one correspondence between the states in a free Fermi gas and
an interacting quantum  system. This allows one to use the
non-interacting language in describing quantum liquids. In particular,
one can regard the interacting Fermi system as a gas of elementary
quasiparticles. In this approach, the number of parameters describing
the state of the system is less than within the exact description.
Thus, an elementary excitation is not a stationary state but a wave
packet of stationary states which spreads with time. This leads to a
finite life-time of elementary excitations $\tau_{\bf p}$ away from
the Fermi surface. However, if the inverse life-time is smaller than
the excitation energy $\tau_{\bf p} \xi_{\bf p} \gg 1 $, one can
regard the excitations as stable particles. The effective mass $m^*$
of these particles is renormalized by the electron-electron
interactions and can be quite different from the non-interacting bare
electron mass (the band mass, $m$). The concept of the electron
effective mass has been a subject of investigation for over fifty
years. Surprisingly, the question of the effective mass temperature
dependence had never been addressed until very
recently.\cite{CM1,CM,Short} This can be partially explained by the fact
that most of the work was performed back in the fifties and sixties,
when only the three-dimensional case was of interest. In typical
three-dimensional systems, the Fermi energy is very high compared to
the temperatures relevant to experiments ({\em i.e.}, in simple metals:
$T_{\rm F} \sim 10^4\,$K) and thus any temperature corrections are
negligible. The Fermi energy in realistic semiconductor-based
two-dimensional systems ({\em e.g.} Si MOS structures, GaAs
heterostructures and quantum wells) may be as low as $1K$, which makes
the issue of the temperature dependence of Fermi liquid parameters
extremely important. In this paper, we obtain analytical results for
the quasiparticle effective mass and the quasiparticle renormalization
factor for two-dimensional and three-dimensional electron systems
interacting via the realistic long-range Coulomb potential.  Our
results employ the standard perturbation theory expansion in the
dynamically screened interaction and are exact in the high-density
limit. We restrict ourselves entirely to the case of an ideal clean
({\em i.e.}, no impurity disorder) and homogeneous electron system with a
parabolic non-interacting energy dispersion.

Before describing the main results and the structure of our paper, let
us briefly discuss previous studies in the subject. The first work
explicitly calculating the effective mass is due to
Gell-Mann,\cite{G-M} who derived a zero temperature correction to the
effective mass due to the Coulomb interaction in the high-density
limit in three dimensions. Galitskii\cite{Gal} has developed a general
scheme of calculating perturbative corrections to the one-particle
spectrum of an interacting Fermi-system. Within this scheme, he
derived corrections to the quasiparticle effective mass and lifetime
for the cases of both a short-ranged interaction and the long-ranged
Coulomb interaction. These studies were again constrained to zero
temperature.  Chaplik\cite{Chaplik} and later Giulianni and
Quinn\cite{GQ} have addressed the issue of the quasiparticle lifetime
temperature dependence having found in two dimensions a non-analytic
contribution to this quantity $E_{\rm F} \tau_{\bf p}^{-1} \propto
{\rm max} \left\{ \xi_{\bf p}^2, T^2 \right\} \ln \left[ {\rm max}
\left\{ \xi_{\bf p}, T \right\}\right] $. This result assures that the
quasiparticles are well-defined excitations as long as $T \ll T_{\rm
  F}$. Very recently, Chubukov and Maslov\cite{CM,CM1} revisited the
problem of non-analytic corrections to the Fermi-liquid theory for the
case of a short-ranged interaction. In particular, they showed that
the leading temperature correction to the effective mass is linear,
similar to the results for the Coulomb interaction case, which had
been reported in a short numerical paper earlier.\cite{Short}

In this  paper, we present detailed purely analytic
calculations of the effective mass renormalization by the Coulomb
interaction. We work within  first order perturbation theory in the
screened interaction, {\em i.e.} within the random-phase approximation
(RPA).  We analytically derive the leading temperature corrections to
the effective mass in the low temperature $T/T_{\rm F} \ll r_{\rm s}$ and
high density $r_{\rm s} \ll 1$ limits. In two dimensions, the leading
correction is positive and linear in temperature with the subleading
term being of the order of $T^2 \ln{T}$ and negative. The linear term
coefficient is found to be a density independent universal number. In
two dimensions, we predict a non-monotonic effective mass temperature
dependence. The point of maximum of the curve $m^*(T)$ is calculated
explicitly and is shown to drift toward higher temperatures as $r_{\rm s}$
increases.

For the sake of completeness, we also calculate $m^*(T)$ for the case
of a three-dimensional electron liquid. At high densities $r_{\rm s}
\ll 1$, the leading contribution is of the order of $T^2
\ln{\left(1/T\right)}$ and positive. As $r_{\rm s}$ increases, the
correction changes its sign and $m^*(T)$ monotonically decreases from
its zero temperature value.

In addition to the quasiparticle effective mass $m^*$, another
important many body Fermi-liquid parameter is the quasiparticle
renormalization factor (the $Z$-factor), which is a measure of the
quasiparticle spectral weight. In particular, the $Z$-factor defines
the size of the effective Fermi surface discontinuity in an
interacting system, and is precisely the size of the discontinuity in
the momentum distribution function $n(p)$. For the non-interacting
Fermi gas, $n(p) = \theta\left( p_{\rm F} - p \right)$, and the
discontinuity is precisely unity whereas for an interacting system
this discontinuity is $Z < 1$.  Note that $Z \ne 0$ implies the
validity of Fermi liquid theory.  To the best of our knowledge, there
has been no consistent microscopic {\em analytic} derivation of the
interaction corrections to the quasiparticle $Z$-factor even in three
dimensions and zero temperature. To fill this gap we calculate
analytically the $Z$-factor. The technical part of this calculation is
found to be more complicated than the effective mass calculation. To
get the correct result, one is required to use the exact forms of the
polarizability to ensure the convergence of the final result.

Our paper is structured as follows: In Sec.~\ref{Sec:1}, we give a
general introduction and derive the basic formulae for the
analytically continued self energy in first order perturbation theory
in the screened interaction in three and two dimensions. In
Sec.~\ref{Sec:1C}, we briefly discuss the structures of the
interaction propagator and the polarization operator in two and three
dimensions. In Sec.~\ref{Sec:3D}, we study the temperature dependence
of the effective mass and also derive the asymptotic formula for the
quasiparticle $Z$-factor in a three dimensional electron liquid.  In
Sec.~\ref{Sec:2D} we study the two-dimensional case and find that the
leading temperature correction to the effective mass is linear and
positive with the subleading term being of the order of $T^2 \ln{T}$ and
negative.  Hence, the effective mass temperature dependence is
non-monotonic. We explicitly derive the temperature $T^*$ at which the
maximum of $m^*(T)$ occurs. We show that the non-analytic contribution
to the effective mass is due to the singularity of the polarization
operator at $\omega \sim v_{\rm F} q$ and $q \ll 1$.  We also discuss
the asymptotic behavior of the quasiparticle renormalization factor in
two dimensions and show that within the RPA approximation the
correction is indeed finite and negative. We emphasize that this result is
due to a subtle cancellation of the logarithmic singularities in the
$\left( {\partial \Sigma \over \partial \varepsilon} \right)$-derivative.  These
kinds of singularities persist in higher orders of  perturbation
theory.

\section{General formulae}\label{Sec:1}

In this section, we give  basic formulae which will be used for
actual calculations. We consider a spin degeneracy factor of $2$
throughout the paper. We denote the quasiparticle effective mass (bare
mass) as $m^*$ ($m$). We use the following small parameters:
$\alpha^{(3D)}=e^2/ {(\pi \hbar v_{\rm F})}$ and $\alpha^{(2D)}=e^2/
{( \hbar v_{\rm F})}$ in three and two dimensions respectively. These
true parameters of the asymptotic expansion are connected with the
usual $r_{\rm s}$-parameter as follows: $r_{\rm s}^{(3D)} = \pi \left( 9 \pi / 4
\right)^{1/3} \alpha^{(3D)}$ and $r_{\rm s}^{(2D)} = \sqrt{2}
\alpha^{(2D)}$. In what follows we will use units $\hbar=k_{\rm B}=1$.


\subsection{Renormalized spectrum}\label{Sec:1A}

The exact Green function for a system of interacting fermions can be
expressed in terms of the self-energy $\Sigma(\varepsilon,{\bf p})$
as follows
\begin{equation}
\label{G-1}
G^{-1}(\varepsilon, {\bf p}) = \varepsilon  -E_0({\bf p})+\mu -
\Sigma(\varepsilon,{\bf p}),
\end{equation}
where $E_0({\bf p}) = p^2 /2m$ is the spectrum of non-interacting
fermions and $\mu \equiv E_{\rm F}$ is the bare chemical
potential.  The Green function can be rewritten as
\begin{equation}
\label{GZ}
G(\varepsilon, {\bf p}) = {Z \over \varepsilon - E({\bf p}) + i
  \gamma(\varepsilon, {\bf p})},
\end{equation}
where $E({\bf p})$ is the renormalized spectrum of excitations,
$\gamma(\varepsilon, {\bf p})$ is the quasi-particle decay rate, and
$Z$ is the residue of the Green function, which determines the jump in
the Fermi distribution at $p=p_{\rm F}$.  In what follows, we mostly will be
 interested in the renormalized single-particle spectrum, which
is connected with the real part of the self-energy as follows:
\begin{equation}
\label{spectrum}
E(p) =  E_0(p)+ {\rm
  Re\,}\Sigma(\varepsilon, \xi_{\bf p})\Biggr|_{\varepsilon=\xi_{\bf p}},
\end{equation}
where we write the self-energy as a function of  a small excitation
energy $\xi_{\bf p}$:
$$
\xi^{(0)}_{\bf p}={p^2 \over 2m} - \mu \approx {p_{\rm F} \over m} \left(p -
p_{\rm F}\right),
$$
$$
\xi_{\bf p}=E(p) - \mu^* \approx {p_{\rm F} \over m^*} \left(p -
p_{\rm F}\right).
$$
The shift of the chemical potential of quasiparticles is determined
by the following equation
\begin{equation}
\label{mu}
\mu^* - \mu =  \Sigma(0,0; \mu^*).
\end{equation}
From the above equations, it follows:
\begin{equation}
\label{xi}
\xi_{\bf p} =\xi^{(0)}_{\bf p} + {\rm Re} \left[ \Sigma\left(\varepsilon=\xi_{\bf p},\xi_{\bf
  p}\right) -
\Sigma\left(0,0\right) \right].
\end{equation}
Solving the linear equation, we obtain the following formula
for the renormalized effective mass:
\begin{equation}
\label{Dyson}
{m^*(T) \over m} = \frac{1}{Z} \left[ 1 + {\partial \over \partial  \xi_{\bf
    p} } \tRe \Sigma(\varepsilon, \xi_{\bf p})
    \right]^{-1}\Biggr|_{\varepsilon,\xi_{\bf p}=0},
\end{equation}
with
\begin{equation}
\label{Z}
Z =  \left[ 1 - {\partial \over \partial \varepsilon}
  \tRe \Sigma(\varepsilon, \xi_{\bf p}) \right]^{-1}\Biggr|_{\varepsilon,\xi_{\bf p}=0}.
\end{equation}

In the perturbative regime, the effective mass reads:
\begin{equation}
\label{mpert}
{m^*(T) \over m} = 1 - \left[  {\partial \over \partial \varepsilon  } + {\partial \over \partial  \xi_{\bf
    p} } \right] \tRe \Sigma(\varepsilon, \xi_{\bf p})
       \Biggr|_{\varepsilon,\xi_{\bf p}=0},
\end{equation}


\subsection{Self-energy in the RPA approximation.}\label{Sec:1B}

In  first order perturbation theory in interaction, the Matsubara
self-energy can be written as (see Fig.~1a):\cite{AGD}
\begin{equation}
\label{Mself}
\Sigma ( \varepsilon_n, {\bf p}) = - T \sum\limits_{\omega_m} {\cal
  G}( \varepsilon_n - \omega_m, {\bf p-q}) {\cal D}(\omega_m, {\bf q}),
\end{equation}
where $ \varepsilon_n = \pi (2 n +1) T$ is the fermion Matsubara
frequency, $\omega_m = 2 \pi m T$ is the boson Matsubara frequency,
and $T$ is the temperature. The function ${\cal D}({\bf q},\omega_m)$
denotes the coupling to a collective mode (phonon, plasmon,
electron-hole excitation, {\em etc.}), {\em i.e.}, an effective
interaction.

For analytical calculations, it is more convenient to use the
self-energy as a function of the real frequency $\varepsilon$ rather
than the Matsubara frequency $\varepsilon_n$.  It is known\cite{Mahan} that in
some cases ({\em e.g.}, when calculating the ground state energy in
the RPA) it is more convenient to do the Matsubara summations first to
avoid divergences arising form the plasmon pole. However, when
calculating the effective mass or the quasiparticle renormalization
factor one can do the analytical continuation first and obtain finite
results. In the calculation of the effective mass in the high-density
limit, the plasmon singularity does not show up in the calculations at
all.\cite{AGD} The calculation of the $Z$-factor is more complicated
but the correct result can be obtained by keeping the exact
$q$-dependence in the polarizability (see Sec.~\ref{Sec:Z3D}). Using
the standard procedure of the analytical continuation, one can obtain
the following expression for the analytically continued
self-energy function:\cite{AGD}
\begin{eqnarray}
\label{Sig_gen}
&&\Sigma_R \left( \varepsilon, \xi_{\bf p} \right) =
-\int {d^d {\bf q} \over (2 \pi)^d}
\int\limits_{-\infty}^{+\infty} {d \omega \over 2 \pi}
 \nonumber \\
&&\times \Biggl\{\tIm G_R
\left(\omega + \varepsilon, {\bf p} - {\bf q} \right)
D_R\left(-\omega, {\bf q} \right) \tanh \left[ { \omega  + \varepsilon \over 2 T} \right] \nonumber \\
&&~~~+ G_R \left( \omega + \varepsilon, {\bf p} - {\bf q} \right)
\tIm D_R\left( \omega, {\bf q} \right) \coth \left[ {\omega
  \over 2 T} \right] \Biggr\},
\end{eqnarray}
where functions labeled with index R are retarded functions, {\em
  i.e.}, functions analytical in the upper half-planes of the complex
frequency and $d$ stands for the dimensionality of space.

\begin{figure}[htbp]
\centering
\includegraphics[width=2.5in]{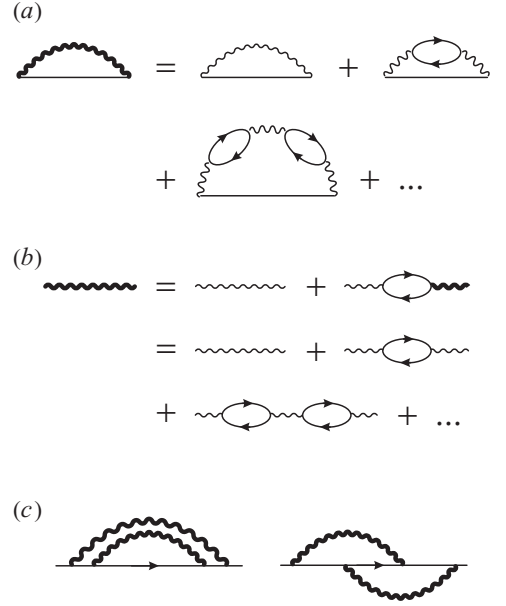}
\caption{(a) The self-energy diagram relevant in the high-density
  limit with the thick wiggly line being the dynamically screened
  Coulomb interaction and the thin wiggly line the bare Coulomb
  interaction. The solid lines correspond to the electron Green
  functions;
(b) The dynamically screened Coulomb interaction infinite series in
  the bare interaction through the bubble diagrams. A bubble
  corresponds to the noninteracting
(``Lindhard'') polarizability.
(c) Examples of higher order diagrams which are negligible in the high
density limit but are important at lower densities.
 }
\label{fig1}
\end{figure}



Within leading order perturbation theory, one can use the bare
electron Green function in Eq.~(\ref{Sig_gen}), which can be written
as
$$
 G_R^{(0)} \left(\varepsilon, {\bf p} \right) = \left[ \varepsilon -
 \xi^{(0)}_{\bf p} + i 0 \right]^{-1}.
$$

If the effective interaction is isotropic (which indeed is the case in
the jellium model we wish to study), the integral(s) over the
directions of ${\bf q}$ can be evaluated and we obtain the following
expressions for the real part of the retarded self-energy function:
\begin{equation}
\label{1+2}
\tRe \Sigma_R \left(\varepsilon, \xi_{\bf p} \right) =
 \Sigma_1 \left( \varepsilon, \xi_p \right)  +  \Sigma_2 \left( \varepsilon, \xi_p \right),
\end{equation}
where in three dimensions:
\begin{equation}
\label{S13D}
\Sigma_1^{(3D)} \left( \varepsilon, \xi_p \right) = \frac{1}{8 \pi^2 v_{\rm F}}
\fint\limits_{-\infty}^{+\infty} d \omega
\int\limits_{q_1(\Omega)}^{q_2(\Omega)}\!\!\! dq q
 \tRe D_R \left( \omega, q \right) \tanh\left[ \frac{\omega
  + \varepsilon}{2 T} \right]
\end{equation}
and
\begin{eqnarray}
\label{S23D}
\nonumber
\Sigma_2^{(3D)} \left( \varepsilon, \xi_p \right) = && \!\!\!\!\!\! - \frac{1}{8 \pi^3
  v_{\rm F}}
\fint\limits_{-\infty}^{+\infty} d \omega \int\limits_{0}^{\infty} dq
q  \ln\left| \frac{\Omega - q^2/{2m} + v_{\rm F} q}{\Omega
  - q^2/{2m} - v_{\rm F} q} \right|\\
&& \!\!\!\!\!\! \times
\tIm D_R \left(  \omega, q \right)
\coth\left[ \frac{\omega}{2 T} \right],
\end{eqnarray}
and in two dimensions
\begin{eqnarray}
\label{S12D}
\nonumber
 \Sigma_1^{(2D)} \left(  \varepsilon, \xi_p \right) =
&& \!\!\!\!\frac{1}{4 \pi^2 v_{\rm F}}
\fint\limits_{-\infty}^{+\infty} d \omega
 \int\limits_{q_1(\Omega)}^{q_2(\Omega)}
\frac{dq}
{\sqrt{1 - \left( \frac{\Omega - q^2/{2m}}{v_{\rm F} q} \right)^2}}\\
&& \!\!\!\!  \times \tRe D_R \left(  \omega, q \right) \tanh\left[ \frac{\omega
  + \varepsilon}{2 T} \right]
\end{eqnarray}
and
\begin{eqnarray}
\label{S22D}
\nonumber
\Sigma_2^{(2D)} \left( \varepsilon, \xi_p \right) = - \frac{1}{4 \pi^2
  v_{\rm F}}
\fint\limits_{-\infty}^{+\infty} d \omega
\left[ \int\limits_0^{q_1(\Omega)}
+ \int\limits_{q_2(\Omega)}^{\infty} \right] \\
\times \frac{dq}
{\sqrt{ \left( \frac{\Omega - q^2/{2m} }{v_{\rm F} q} \right)^2 - 1}}
\tIm D_R \left( \omega, q \right)
\coth\left[ \frac{\omega}{2 T} \right].
\end{eqnarray}
In Eqs.~(\ref{S13D}), (\ref{S23D}), (\ref{S12D}), and (\ref{S22D}), we
introduced the following notations for the sake of brevity:
$$
\Omega = \omega + \varepsilon - \xi_p
$$
and $q_2\left( \Omega \right) \geq q_1\left( \Omega \right) \geq 0$
which are the solutions of the equation $\left| \Omega - q^2/{2m} \right| = v_{\rm F} q$.

\subsection{Effective interaction and the polarization operator.}\label{Sec:1C}

The appropriate propagator in the case of an electron liquid with
long-range Coulombic forces between electrons is given by the sum of
the ladder of bubble diagrams and has the typical RPA form (see Fig.~1b):
\begin{equation}
\label{D(q,w)}
 D( \omega, q) = {V (q)  \over 1 + V (q) \Pi
  (\omega, q) },
\end{equation}
where $V(q)$ is the bare Coulomb interaction [$V(q)=4 \pi e^2 /q^2$ in
three dimensions and $V(q)=2 \pi e^2 /q$ in two dimensions] and $\Pi
(\omega,q)$ is the polarizability bubble:
\begin{equation}
\label{Polar}
\Pi \left(\omega, {\bf q} \right) =  2
\int {d \varepsilon \, d ^d {\bf p}  \over \left(2 \pi \right)^{(1+d)}}
G^{(0)} \left( \varepsilon, {\bf p} \right)  G^{(0)} \left( {\varepsilon + \omega}, {\bf p + q}\right).
\end{equation}
At zero temperature, the polarizability was calculated by Lindhard\cite{Lindhard} and
Stern\cite{Stern} in three and two dimensions respectively. We will need the
exact expressions, which can be conveniently written in terms of the
dimensionless parameters $u = \omega/\left( v_{\rm F} q \right)$ and
$x = q/ (2 p_{\rm F})$. In three dimensions it reads
\begin{eqnarray}
\label{ReP3}
\nonumber
\tRe \Pi^{(3D)} \left( u, x \right) = \frac{\nu^{(3D)}}{2} \Biggl\{ &&\!\!\!\!\!\! 1 + {1 \over 4x}
\left[   1 -\left( x + u \right)^2 \right] \ln \left|
\frac{1+x+u}{1-x-u} \right| \\
&& \!\!\!\!\!\!\!\!\!\!\! + {1 \over 4x}
\left[  1 -\left( x - u \right)^2 \right] \ln \left|
\frac{1+x-u}{1-x+u} \right| \Biggr\}
\end{eqnarray}
and the imaginary part for $u>0$
[$\tIm \Pi \left( u, x \right) = - \tIm \Pi \left( -u, x \right)$]:
\begin{eqnarray}
\label{ImP3}
\nonumber
\tIm \Pi^{(3D)} \left( u, x \right) = -\frac{\pi \nu^{(3D)}}{2} \Biggl\{  u
  \,\, \theta\left( \left| 1 - x \right| - u \right) \\
+ \frac{1}{x} \left[ 1 - \left( x  -u \right)^2 \right]
\theta\left( \left| 1 + x\right| - u \right)
\theta\left( u -  \left| 1 - x\right| \right) \Biggr\},
\end{eqnarray}
where $\nu^{(3D)} = m p /\pi^2$ is the density of states at the Fermi
surface.

In two dimensions the polarizability has the following explicit form:
\begin{eqnarray}
\label{ReP2}
\nonumber
\tRe \Pi^{(2D)} \left( u, x \right) = \nu^{(2D)} \tRe \Biggl\{  1 - {{\rm
    sgn\,} \left( x+u \right) \over 2x}
\sqrt{ \left( x +  u \right)^2 -1} \\
\!\!\!\!-
{{\rm
    sgn\,} \left( x - u \right) \over 2x}
\sqrt{ \left( x - u \right)^2 -1}
\Biggr\}
\end{eqnarray}
and the imaginary part for any $u$:
\begin{eqnarray}
\label{ImP2}
\!\! \tIm \Pi^{(2D)}\! \left( u, x \right) =\!\! \frac{\nu^{(2D)}}{2x}
\tRe \Biggl\{
\!\!\! \sqrt{1 - \left( x + u \right)^2} - \!\! \sqrt{1 - \left( x - u \right)^2} \Biggr\},
\end{eqnarray}
with  $\nu^{(2D)} = m  /\pi$ being the two-dimensional density of states
at the Fermi line.

Let us emphasize that both three dimensional and two dimensional
polarizabilities are non-analytic functions at $\left| u \pm x \right|
=1$.  Usually this singularity is associated with Friedel oscillations
and Kohn-Luttinger effect,\cite{KL} {\em i.e.} with the famous Kohn anomaly
at $q = 2 p_{\rm F}$ and $\omega \to 0$. In the case of a dense
Coulomb liquid, the typical momenta are small $x = q/ (2 p_{\rm F})
\ll 1$, but $u=\omega / (v_{\rm F} q)$ can be of the order of unity or
even larger. As we shall see, exactly this domain of parameters
({\em i.e.}, small momentum transfers) is
responsible for non-analytic contributions to the effective mass
temperature dependence and the quasiparticle $Z$-factor.
The usual Kohn singularity ($x=1$) in the static polarizability also
gives rise to a non-analytic temperature dependence but this effect is
parametrically smaller than the dynamic screening effects in the limit
$r_{\rm s} \ll 1$.

The issue of the temperature dependence of the polarization operator
was recently reconsidered in great details by Chubukov and
Maslov\cite{CM} (see also Refs.~[\onlinecite{Stern80}] and
[\onlinecite{SDS}]) who found that in the vicinity of the Kohn
singularity the polarizability has a linear $T$ correction, which is
important in the case of a short range interaction case. In the case
of the long-range Coulombic forces, the $q$-dependence of the
propagator (\ref{D(q,w)}) becomes crucial and in the high-density
limit the results are determined by the region $x =q/(2 p_{\rm F})
\sim r_{\rm s} \ll 1$ in two dimensions ($x \sim \sqrt{r_{\rm s}}$ in
three dimensions).  Hence, in the leading order in $r_{\rm s}$, only
the region $x \ll 1$ is important.  In this region, the leading
temperature correction to the polarizability is of the order of $T^2$
in any dimensionality.  As we shall see below, the leading order
temperature corrections to the effective mass and the $Z$-factor are
parametrically larger than $T^2$.  Therefore, in the high-density
limit, the temperature corrections to the polarization bubble give
negligible contributions to the quasiparticle spectrum and in the case
$T / E_{\rm F} \ll r_{\rm s} \ll 1$, one can use the zero temperature
results for the polarizability [see Eqs.~(\ref{ReP3}), (\ref{ImP3}),
(\ref{ReP2}), and (\ref{ImP2})] in the calculations of the effective
mass and the $Z$-factor.

\subsection{Collective modes}

The usual practice is to expand the polarizability functions in
$x \ll 1$; in which case the polarizability becomes the function of
just one variable $u = \omega/(v_{\rm F} q)$. The limit $u \ll 1$
corresponds to the electron-hole branch of excitations. The
corresponding retarded electron-hole propagator is (in three
dimensions)
\begin{equation}
\label{Deh3}
D_{\rm eh}^{(3D)}(\omega,q) = 4 \pi e^2 \left[q^2 + \varkappa_3^2 - i
{\pi \over 2} \varkappa_3^2 \left({\omega \over v_{\rm F}q} \right)  \right]^{-1},
\end{equation}
where $\varkappa_3  = 2 p_F \sqrt{\alpha^{(2D)}}$ is the inverse
screening length.

In two dimensions the electron-hole propagator reads
\begin{equation}
\label{Deh2}
D_{\rm eh}^{(2D)}(\omega,q) = 2 \pi e^2 \left[ q + \varkappa_2 - i
\varkappa_2 \left( {\omega \over v_{\rm F} q} \right)  \right]^{-1},
\end{equation}
where $\varkappa_2  = 2 p_F \alpha^{(2D)}$ is the inverse screening
length in two dimensions.

The opposite limit $u \gg 1$ corresponds to the plasmon branch.
The spectrum of plasma waves is determined by the equation
$$
1 + V(q)\Pi(\omega,q) = 0.
$$
In three dimensions the spectrum has the following well-known form
$\omega_{\rm pl}^{(3D)}(q) = \sqrt{ \omega_0^2 + {3 \over 5} v_{\rm
    F}^2 q^2}$, with $\omega_0 = 4 \pi e^2 n/m$. In two dimensions, the
plasmons are gapless $\omega_{\rm pl}^{(2D)}(q) = \sqrt{aq + bq^2}$
with $a = 2 e^2 E_{\rm F}$ and $b = {3 E_{\rm F} / (2 m)}$.

Within the first approximation in the interaction, the imaginary part
of the polarization operator at $\omega \gg v_{\rm F} q$ is zero.
This is exactly the source of the well-known plasmon singularity in
self-energy calculations.  However, higher order diagrams deliver non-zero
contributions to the imaginary part. Taking into account this fact, we
can write down the following expression for the retarded plasmon
propagator:
\begin{eqnarray}
\label{Dpl}
D_{\rm pl} (\omega,q) = {1 \over 2} V(q) \omega
\Big[&&\!\!\!\!\! {1 \over \omega - \omega_{\rm pl}(q) + i 0} \nonumber \\
&& \!\!\!\!\!  + {1 \over \omega + \omega_{\rm pl}(q) + i 0} \Big]
\theta\left(q_{\rm m} - q \right),
\end{eqnarray}
where $q_{\rm m}$ is the wave-vector at which the strong Landau
damping commences.

We shall see that the leading temperature correction to the effective mass
both in two and three dimensions comes mostly from the region of $u \sim 1$,
which is neither plasmon nor electron-hole region. In actual
calculations, we do not separate the screened Coulomb propagator into
the electron-hole and plasmon branches. Moreover, for the calculation of
the renormalization factor one is required  to keep the exact
polarizability function (without expanding on $x \to 0$).

\section{Three-dimensional case} \label{Sec:3D}

In this section we present analytic calculations of the effective mass
and the quasiparticle $Z$-factor in a three-dimensional dense electron
liquid. Throughout this section, $\nu = m p /\pi^2$ is the
three-dimensional density of states and $\alpha = e^2/ \left( \pi
\hbar v_{\rm F} \right)$ is the appropriate expansion parameter. The
main results are Eqs.~\eqref{mass3} and \eqref{Z3} given below.

\subsection{Effective mass}\label{Sec:m3D}

In the high density limit, the correction to the effective mass is
determined by the on-shell equation \eqref{mpert}, which means that we
can put $\varepsilon = \xi_p$ and study the self-energy as a function
of just one variable [see Eqs.~\eqref{ReP3}]. On the shell, we have
the following expression:
\begin{eqnarray}
\label{on-shell3}
{\partial  \over \partial \xi}  \tRe \Sigma \left
( \epsilon=\xi, \xi \right) \Biggr|_{\xi = 0} = &&  \frac{1}{8 \pi^2 v_{\rm F}}
\fint\limits_{-\infty}^{+\infty} {d \omega \over 2 T} {1 \over \cosh^2
  \left[ {\omega \over 2 T} \right]} \nonumber \\
&&   \times \int\limits_{q_1[\omega - E_0(q) ]}^{q_2[\omega - E_0(q)]}\!\!\! dq q
 \tRe D \left( \omega,q \right).
\end{eqnarray}

It is convenient to separate the static and dynamic propagators:
\begin{equation}
\label{st+dyn}
D(\omega,q) = D_{\rm st}(q) +  D_{\rm dyn}(\omega,q) \equiv
D(0,q) + \left[ D(\omega,q) - D(0,q) \right].
\end{equation}
 The static propagator has the form:
\begin{equation}
\label{static3}
D_{\rm st}(x) = {\alpha \over \nu} \, {1 \over x^2 + \alpha}.
\end{equation}

At low temperatures, frequencies $\omega$ in the integral
\eqref{on-shell3} are of the order of temperature or even lower. We
consider the following asymptotic regime of ultra-low temperatures:
$$
\delta = \omega / (4 E_{\rm F}) \sim T/E_{\rm F} \ll \alpha \ll 1.
$$
In this limit, the real part of the dynamic propagator has following form [see
Eq.~\eqref{ReP3}, \eqref{ImP3}]:
\begin{eqnarray}
\label{dyn3}
\nonumber
\tRe D_{\rm dyn}(\delta,x) = && \!\!\!\!\!\!\! - {\alpha \over \nu} \,
\left( { \pi \alpha \delta \over 2} \right)^2  \left( x^2 + \alpha
\right)^{-1}\\
&& \!\!\!\!\!\!\! \times \left[ x^2 \left( x^2 + \alpha
\right)^2 +  \left( { \pi \alpha \delta \over 2} \right)^2 \right]^{-1}.
\end{eqnarray}
Using Eq.~\eqref{on-shell3} and propagators \eqref{static3} and
\eqref{dyn3}, one can calculate corrections to the effective mass
(expanding on the small parameter $\delta \sim T/E_{\rm F}$) and see
that the static contribution, indeed, solely determines the
renormalization of the effective mass at zero temperature.  However,
it gives temperature corrections of the order of $T^2$ only. The
dynamic part gives zero contribution to the zero temperature effective
mass, but gives parametrically larger temperature dependent
correction.  The final result reads:
\begin{equation}
\label{mass3}
{m^*\left(T\right) - m  \over m} = - {\alpha \over 2} \ln{1 \over
  \alpha} + {\pi^2 \over 96} \left( T \over E_{\rm F} \right)^2 \ln{E_{\rm F}
  \over T}.
\end{equation}
The first term indeed coincides with the old result of Gell-Mann.  The
non-analytic temperature correction given in the second term of
Eq.~\eqref{mass3} is a new result.  Let us emphasize that the leading
correction is positive only in the high-density limit. In
Ref.~[\onlinecite{Short}], it was shown that the leading $T^2 \ln{T}$
correction changes sign at lower densities ($\alpha^{(3D)} \sim 1$).
The density dependence comes from large momentum transfers $q$, in
particular from the vicinity of the $2 p_{\rm F}$-anomaly, which
becomes increasingly important at lower densities.

\subsection{$Z$-factor}\label{Sec:Z3D}

Unlike in the effective mass calculation of the preceding section, the
quasiparticle $Z$-factor can not be calculated within the on-shell
method, and therefore the problem is more complicated.  In particular,
one has to consider both contributions to the self-energy given in
\eqref{S13D} and \eqref{S23D}. The first contribution at zero
temperature reads:
\begin{eqnarray}
\label{S1Z3}
{\partial  \over \partial \varepsilon}  \Sigma_1 \left(
\varepsilon, \xi =0 \right) \Biggr|_{\varepsilon = 0} =
{\nu \over 2} \int\limits_0^{\infty} dx x
\fint\limits_{-\infty}^{\infty} du
 \tRe D \left( u,x \right) \nonumber \\
\times {\partial \over \partial u}
\left[ \theta \left( 1 - \left| u - x \right| \right)
{\rm sgn \,} u  \right].
\end{eqnarray}
In the three-dimensional case, the integral over $u$ reduces to a $\delta$-function integration and we
have:
\begin{eqnarray}
\label{S1Z3-2}
{\partial  \over \partial \varepsilon}  \Sigma_1 \left(
\varepsilon, \xi =0 \right) \Biggr|_{\varepsilon = 0} \!\!\!\!\!\! =
\nu  \int\limits_0^{1} dx x
\Biggl[&& \!\!\!\!\!   D \left( 0, x \right) - {1 \over 2}  \tRe D \left(
1+x, x \right)  \nonumber \\
&& \!\!\!\!\!\!\!\!\! - {1 \over 2}  \tRe D \left(  1-x, x \right) \Biggr].
\end{eqnarray}


From Eq.~\eqref{S23D}, we derive the second contribution:
\begin{eqnarray}
\label{S2Z3}
{\partial  \over \partial \varepsilon}  \Sigma_2 \left(
\varepsilon, \xi =0 \right) \Biggr|_{\varepsilon = 0} \!\!\!\!\!\! = &&
\!\!\!\!\!\! {\nu \over \pi}  \int\limits_0^{\infty} dx \,  x \, \int\limits_0^{\infty}
  \tIm D \left(u, x \right) \nonumber \\
&& \!\!\!\!\!\!  \times \left[ {1 \over 1 - \left( u - x \right)^2}
+ {1 \over 1 - \left( u + x \right)^2} \right].
\end{eqnarray}
Using Eqs.~(\ref{ReP3}), (\ref{ImP3}), and (\ref{Dpl}),
one can evaluate integrals (\ref{S1Z3-2}) and (\ref{S2Z3}) and obtain
\begin{eqnarray}
\label{res3Z2}
{\partial  \over \partial \varepsilon}  \Sigma \left(
\varepsilon, \xi =0 \right) \Biggr|_{\varepsilon = 0} \!\!\!\!\!\! =
- {\alpha \over \pi} \int\limits_0^{\infty} {du \over 1 + u^2}
\ln\left[ {1 \over 1 -
  u \, {\rm atan}\, {1 \over u} } \right].
\end{eqnarray}
Evaluating the remaining integral numerically, we derive the final
result for the quasiparticle renormalization factor:
\begin{equation}
\label{Z3}
Z = 1 - 1.067 \alpha.
\end{equation}
This   asymptotic result is in a very good agreement with numerical
simulations.\cite{Mahan}


\section{Two-dimensional case} \label{Sec:2D}

In this section we present analytic calculations of the effective mass
and quasiparticle $Z$-factor in a two-dimensional dense electron
liquid.  Throughout this section, $\nu = m /\pi$ is the two
dimensional density of states and $\alpha = e^2/ \left( \hbar v_{\rm
  F} \right) \ll 1$ is the appropriate expansion parameter. The main
results are Eqs.~\eqref{mass2}, \eqref{T*}, and \eqref{Z2} given below.

\subsection{Effective mass}\label{Sec:m2D}

The calculation of the effective mass in two dimensions is analogous
to the calculation in three dimensions (see Sec.~\ref{Sec:Z3D}). The
only difference is the square root function, which appears in all
two-dimensional expressions [see, {\em e.g.}, Eqs.~\eqref{S12D} and
~\eqref{S22D}]. This square root singularity is identical to the
singularity in the Stern's polarizability function given in
Eqs.~\eqref{ReP2} and \eqref{ImP2}. The combination of these two
singularities leads to a stronger temperature dependence (linear as we
shall see) as compared to the three-dimensional case. From
Eqs.~\eqref{mpert} and ~\eqref{S12D}, we get the following on-shell
expression for the two-dimensional electron effective mass:
\begin{eqnarray}
\label{dm2}
{m^* - m \over m} =
-{ \nu \over 2  \pi}
\fint\limits_{-\infty}^{\infty} {d \omega \over 2 T}
{1 \over \cosh^2 {\omega \over 2 T} } I(\omega),
\end{eqnarray}
where we have introduced the following integral:
\begin{equation}
\label{I}
I(\omega) = \int\limits_{x_1(\omega)}^{x_2(\omega)}
dx
{x \tRe D \left(  \omega, x \right) \over \sqrt{\left[ x^2 - x_1^2(\omega) \right]
\left[ x_2^2(\omega) - x^2 \right]}},
\end{equation}
where $x = q/\left( 2 p_{\rm F} \right)$ and $x_{1,2}$ are the  solutions of the
equation $\left| {\omega \over 4 E_{\rm F} x} -x \right| = 1$. In the limit
of low temperatures: $x_1 \approx \left|\delta \right| = \left| \omega \right| / (4 E_{\rm F})$ and
$x_2 \approx 1$.

Again we re-write the propagator as a sum of  static and
dynamic terms \eqref{st+dyn}. The two-dimensional static propagator has the form
\begin{equation}
\label{Dst2}
D_{\rm st} (x) = {\alpha \over \nu} {1 \over x + \alpha}.
\end{equation}
The real part of the dynamic propagator in the limit $|\omega| \sim T
\ll E_{\rm F}$ reads:
\begin{eqnarray}
\label{dyn2}
\tRe D_{\rm dyn}(\delta, x) = && \!\!\!\!\!\!\! - {\alpha \over \nu} \,
\left( \alpha \tIm \Pi \right)^2 \left( x + \alpha
\right)^{-1}\\
&& \!\!\!\!\!\!\! \times \left[  \left( x + \alpha
\right)^2 +  \left( \alpha \tIm \Pi \right)^2 \right]^{-1},
\end{eqnarray}
where $\tIm \Pi$ is defined by Eq.~\eqref{ImP2}.

Expanding in the small parameter $\delta$, we obtain
the contribution to the integral \eqref{I} due to the static
propagator
$$
I_{\rm st} (\delta) = {\alpha \over \nu} \left[ \ln{1 \over \alpha} - {\delta^2 \over 2
  \alpha^2} \ln{1 \over \left| \delta \right|} \right]
$$
and the ``dynamic part''
$$
I_{\rm dyn} (\delta) =  {\alpha \over \nu} \left[ -{\pi \over 2} {\left| \delta \right| \over
  \alpha} +  {3
  \delta^2 \over \alpha^2} \ln {1   \over \left| \delta \right|} \right].
$$
Evaluating the elementary integral, we obtain the final result for
the temperature dependent effective mass in the second leading order in temperature:
\begin{equation}
\label{mass2}
{m^*\left(T\right) - m  \over m} = - {\alpha \over \pi} \ln{1 \over
  \alpha} + {\ln{2} \over 4} \left( T \over E_{\rm F} \right)  - {5 \pi
  \over 48 \alpha} \left( T \over E_{\rm F} \right)^2 \ln{E_{\rm F}
  \over T}.
\end{equation}
Let us emphasize that Eq.~\eqref{mass2} is valid in the low
temperature and high-density limit: $T/E_{\rm F} \ll \alpha \ll 1$ and
only for subthermal particles: $\varepsilon \ll T$.  We see that the
leading term is linear in temperature and the coefficient is a
universal density independent number. This universal behavior is true
only in the high density limit. There are other linear-$T$
contributions, such as the one due to the temperature dependence of
the polarizability in the vicinity of the Kohn singularity (considered
in the paper of Chubukov and Maslov \cite{CM} for short-range
interactions).  In the case of the long-range Coulomb interaction, the
Kohn anomaly leads to a linear-$T$ term proportional to the Coulomb
expansion parameter $\alpha$. Similar $\alpha$-dependence of the
linear slope was discovered in RPA numerical calculations in
Ref.~[\onlinecite{Short}]. In the high-density limit, this density
dependent linear-$T$ term is asymptotically smaller than the main
universal contribution [the second term in Eq.~\eqref{mass2}] and
therefore not shown in Eq.~\eqref{mass2}.

\begin{figure}[htbp]
\centering
\includegraphics[width=3in]{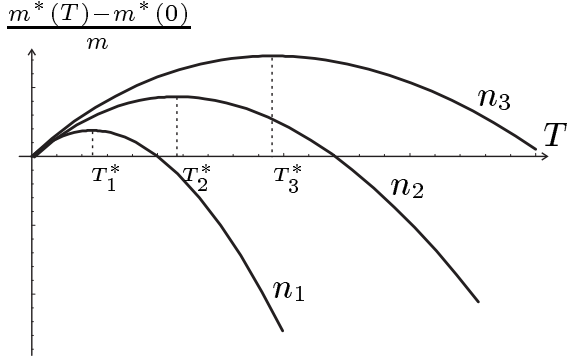}
\caption{Asymptotic effective mass temperature dependence [see
  Eq.~\protect{\eqref{mass2}}] for three different densities $n_1 >
 n_2 > n_3$. The slope of the curves is universal at high densities
[$\sqrt{n}/(m e^2 ) \gg 1$].
 }
\label{fig2}
\end{figure}

From Eq.~\eqref{mass2}, we see that the effective mass temperature
dependence is non-monotonic. A maximum occurs at a temperature $T^*$,
which within the logarithmic accuracy has the form:
\begin{equation}
\label{T*}
{T^* \over E_{\rm F}} = {6 \ln{2} \over 5 \pi}\, {\alpha \over \ln{1 \over
    \alpha}} \approx 0.26\,  {\alpha \over \ln{1 \over
    \alpha}} \ll \alpha.
\end{equation}
This result is formally within the limits of applicability of our
theory.  We see that the point of maximum of the curve $m^*(T)$
drifts toward higher temperatures as the density decreases. This
tendency is preserved at lower densities as well (within the RPA
approach). Such a maximum in $m^*(T)$ and a density dependent
$T^*$ were also discovered in our recent numerical
calculation.\cite{Short} \vspace*{-0.1in}
\subsection{$Z$-factor}\label{Sec:Z2D}

The analytical calculation of the $Z$-factor in two dimensions is
technically a very demanding problem. The mixture of two
singularities, the Kohn singularity in the polarizability and the
identical square root singularity in Eqs.~\eqref{S12D} and
\eqref{S22D} arising from the two-dimensional phase space, leads to a
complicated structure of the integrals in Eqs.~\eqref{S12D} and
\eqref{S22D}, each being a truly divergent quantity. The logarithmic
divergence gets cancelled (at least within the RPA), but to see this
cancellation one is required to keep the exact $x$ and $u$ dependences
in the Stern's polarizability function. Moreover, the technical method used in the
three-dimensional calculation of the $Z$-factor [see
Sec.~\ref{Sec:Z3D}, Eq.~\eqref{S1Z3}] is not applicable here because
of the square-root singularity.

Let us now study the two-dimensional $Z$-factor in more details. The
quasiparticle renormalization factor is determined by the energy
derivative of the self-energy. The latter can be written as a sum of
two terms [see Eqs.~\eqref{S12D} and \eqref{S22D}]:
\begin{eqnarray}
\label{S1Z2}
{\partial  \over \partial \varepsilon}  \Sigma_1 \left(
\varepsilon, \xi =0 \right) \Biggr|_{\varepsilon \to 0}  \!\!\!\! =&& \!\!\!\!
-{\nu \over 2 \pi} \int\limits_0^{\infty} dx x
\int\limits_{0}^{\infty} du
\left[ {\partial \over \partial u} \tRe D \left( u, x \right) \right] \nonumber \\
&&  \!\!\!\! \!\!\!\! \!\!\!\! \times \left[
{\theta \left( 1 - \left| u - x \right| \right) \over \sqrt{1 - \left( u - x \right)^2}} +
{\theta \left( 1 - \left| u + x \right| \right) \over \sqrt{1 - \left( u + x \right)^2}}
\right]
\end{eqnarray}
and
\begin{eqnarray}
\label{S2Z2}
{\partial  \over \partial \varepsilon}  \Sigma_2 \left(
\varepsilon, \xi =0 \right) \Biggr|_{\varepsilon \to 0} =
{\nu \over 2 \pi} \int\limits_0^{\infty} dx x
\int\limits_{0}^{\infty} du
\left[ {\partial \over \partial u} \tIm D \left(  u, x \right) \right] \nonumber \\
\times \left[
{\theta \left( \left| u - x \right| -1 \right) \over \sqrt{ \left( u -
    x \right)^2 -1 }} -
{\theta \left( \left| u + x \right| -1 \right) \over \sqrt{\left( u +
    x \right)^2 - 1}}
\right].
\end{eqnarray}
The frequency dependence (hence, the $u$-dependence) of the propagator
$D(u,x)$ is due to the polarizability $\Pi(u,x)$, which contains
exactly the same square root functions as the ones in
Eqs.~\eqref{S1Z2} and \eqref{S2Z2}. This leads to a logarithmic
divergence of each of the above integrals at $u = 1 \pm x$. The
``singular'' contributions have the following forms:
\begin{eqnarray}
\label{S1sing}
{\partial  \over \partial \varepsilon}  \Sigma_1^{\rm (sing)}
\left(
\varepsilon, \xi = 0 \right) \Biggr|_{\varepsilon \to 0}   =
{1 \over  \pi} \int\limits_0^{\infty} dx x
\int\limits_{0}^{1+x} du
{\left[  \tRe D \left( u, x \right) \right]^2 \over \sqrt{1 - \left( u
    - x \right)^2}} \nonumber \\
\times {\partial  \tIm \Pi \over \partial u}  \tIm \Pi\,
\left[ x + {\alpha \over \nu} \tRe \Pi \right]
\end{eqnarray}
and
\begin{eqnarray}
\label{S2sing}
{\partial  \over \partial \varepsilon}  \Sigma_2^{\rm (sing)}
\left(
\varepsilon, \xi = 0 \right) \Biggr|_{\varepsilon \to 0}   =
{1 \over  \pi} \int\limits_0^{\infty} dx x
\int\limits_{1-x}^0 du
{\left[  \tRe D \left( u, x \right) \right]^2 \over \sqrt{ \left( u
    + x \right)^2 - 1}} \nonumber \\
\times {\partial  \tRe \Pi \over \partial u}  \tIm \Pi\,
\left[ x + {\alpha \over \nu} \tRe \Pi \right].
\end{eqnarray}

Each of these quantities is logarithmically divergent $ \left(
{\partial \Sigma_{1,2} \over \partial \varepsilon} \right) \sim \pm
\alpha \ln{\varepsilon} \to \infty$.  We emphasize that in two
dimensions the real and imaginary parts of the polarizability have
almost identical analytic structures, in contrast to the three
dimensional case in which the imaginary and real parts are basically
independent functions with quite different properties.  Using
Eqs.~\eqref{S1sing} and \eqref{S2sing}, one can check that this ``symmetry''
of the two-dimensional polarizability leads to an exact cancellation
of the logarithmic divergence and to a finite result.

Let us emphasize that this kind of dangerous singularities appear in
any order of the perturbation theory in interaction (see, {\em e.g.},
Fig.~1c). It is not {\em a priori} obvious how (and if) the
singularity, which is cut-off only by temperature or energy
$\varepsilon$, is cancelled in higher order diagrams. We do not have a
general argument for why the divergence must cancel in each order, but
we do know that they cancel to this order. It is essential to clarify
this point to assure that the quasiparticle $Z$-factor does not vanish
logarithmically $Z^{-1}(\varepsilon) \sim \ln{\varepsilon}$ and to
make certain that the usual Landau Fermi liquid theory is preserved in
two dimensions.  This issue is currently being studied by us. We
believe that the Fermi liquid theory is preserved but it needs to be
demonstrated explicitly.

Within the RPA, the zero temperature $Z$-factor can be proven to be
finite.\cite{SDSZ} One can formally define the quasiparticle
$Z$-factor at finite temperatures via relation \eqref{Z}. Studying the
leading temperature correction to the energy derivative of the
self-energy is quite similar to the case of the on-shell derivative
case. The leading term can be shown to be linear and negative:\cite{Loss}
\begin{equation}
\label{Z2}
Z(\alpha, T) \approx 1 - \left( {1 \over 2} + {1 \over \pi} \right) \alpha - c \, \left( {T \over E_{\rm
    F}} \right),
\end{equation}
where $c$ is a constant of the order of unity.

\vspace*{-0.1in}
\section{Conclusion}\label{Sec:Conclusion}
In this work we have developed the analytic leading order theory
for the temperature dependent quasiparticle effective mass,
$m^*(T)$, and the quasiparticle renormalization factor for two-
and three-dimensional interacting electron systems. Our results
are asymptotically exact in the low temperature high-density
limits for the case of the realistic long-range Coulomb
interaction, and thus we are complementary to the recent
analytical work of Ref.~[\onlinecite{CM}], which considers a
short-range repulsive interaction. It is interesting to note that
$m^*(T)$ has an unexpected linear-$T$ correction (rather than
$T^2$) both in our theory and in the theory of Chubukov and
Maslov; but in our case the correction is positive opposite to the
short range case. This immediately leads to the conclusion that
the leading correction to $C_V/T$ (where $C_V$ is the specific
heat) is not universal --- our long range interaction produces a
positive linear-$T$ term in the leading order in contrast to the
negative sign obtained in Ref.~[\onlinecite{CM}]. This unexpected
linear-$T$ term appears due to the non-analyticity of the
polarizability function. This non-analyticity has potentially
important consequences for quantum critical phenomena as discussed
recently in Ref.~[\onlinecite{Pepin}].

Our analytic results are also in  agreement with recent
numerical studies of the temperature dependent effective
mass,\cite{Short} which employed the random-phase approximation at
lower densities. It was shown that the linear-$T$ correction persists
at lower densities as well (at least within the RPA) and the
qualitative behavior of the effective mass remains the same with the
only difference being the density dependent slope of the $m^*(T)$
curve at $T \to 0$ and $r_{\rm s} > 1$ (for high densities, it was shown to
be density independent in agreement with the analytical results
reported in the present paper). As the RPA  is believed to be
qualitatively reliable at lower densities as well, we expect our
results to be quite general and qualitatively applicable to realistic
two-dimensional electron systems. We also should point out that
although the temperature dependent effective mass renormalization has
only been calculated numerically very recently,\cite{Short} there is a
vast literature of  numerical studies of zero temperature many-body
effects in the three-dimensional and two-dimensional interacting electron systems.
We cite in this context only two rather comprehensive references:
Ref.~[\onlinecite{Hedin}] for  three-dimensional systems and
Ref.~[\onlinecite{SDSZ}] for two-dimensional systems.

We have proved that the quasiparticle $Z$-factor in two dimensions
is finite at least within the random-phase approximation. However,
we would like to emphasize that each order in perturbation theory
does contain a dangerous logarithmic singularity in this quantity.
It is known that in two dimensions higher order diagrams may
contain important effects (see, {\em e.g.},
Ref.~[\onlinecite{ChubKL}], in  which it was shown that the static
Kohn-Luttinger effect\cite{KL} in two dimensions is ``hidden'' in
third order perturbation theory only). It is therefore essential
to prove that the cancellation of the logarithmic singularity
(which would otherwise lead to a logarithmically vanishing
$Z$-factor and to a marginal Fermi liquid\cite{LV}) takes place in
higher orders. This important question will be considered
elsewhere. \vspace*{-0.2in}
\begin{acknowledgments}
\vspace*{-0.1in} This work was supported by the US-ONR, LPS, and
DARPA. The authors are grateful to Andrei Chubukov for valuable
discussions.
\end{acknowledgments}
\bibliography{mass}

\end{document}